# PHOTO-EMISSION PROPERTIES OF QUASI-ONE-DIMENSIONAL CONDUCTORS

**Ž. Agić, P. Županović**

*Department of Physics, Faculty of Science and Art, University of Split, Split, Croatia*

**A. Bjeliš**

*Department of Physics, Faculty of Science, University of Zagreb, Zagreb, Croatia*

**Abstract.** We calculate the self-energy of the one-dimensional electron band with the three-dimensional long-range Coulomb interaction within the random phase approximation, paying particular attention to the contribution coming from the electron scatterings on the collective plasmon mode. It is shown that the spectral density has a form of wide feature at the frequency scale of the plasmon frequency, without the presence of quasi-particle δ-peaks. The relevance of this result with respect to experimental findings and to the theory of Luttinger liquids is discussed.

## 1. GENERAL SCOPE

The primary interest into quasi one-dimensional conductors in last decades was focused on their low temperature properties, i.e. on the appearance of density wave and/or superconducting ground states with a variety of accompanying novel collective phenomena [1]. The low temperature regime is usually identified as that in which the coherent deconfinement of electrons due to a finite inter-chain hopping is realized, so that the whole subsystem of conducting electrons (i. e. one or more partially filled bands) behaves as an anisotropic Fermi liquid. At higher temperatures the electronic states are due to the thermalization effectively one-dimensional, and the conducting electrons would have to show characteristics of Luttinger liquids.

The advanced experimental techniques in the last decade, in particular the measurements of angle resolved photoemission spectra (ARPES) [2-5], gave indeed a lot of evidences that the electronic spectral properties in numerous organic and inorganic materials deviate from those of conventional Fermi liquid systems, but also indicated that some observed characteristics are not in agreement with predictions of standard Luttinger scheme. Two main observations having the signature of Luttinger liquid are the absence of quasi-particle δ-peaks in the spectral density $A(\mathbf{k}, \omega)$, and the disappearance of the density of states at the Fermi level ($\omega=0$), given by the power law $N(\omega) \sim \omega^\alpha$, where α, a non-universal *anomalous dimension,* depends on the coupling parameters of the electron-electron interaction [6]. This is to be contrasted with the smooth variation of $N(\omega)$ around the constant value of $N(0)$ in conventional Fermi liquids. The experimental values of α are of the order of unity [2-4], far above the range of values $\alpha \leq 1/8$ for the Tomonaga-Luttinger model with predominant local (on-site Hubbard) interactions. This discrepancy may be resolved by allowing for strong enough site-site contributions to the electron-electron interaction, or by taking fully into account, again presumably strong, long-range Coulomb forces [6,7] (with an appropriate inclusion of three-dimensional screening [8,9]).

The experimental findings that do not suit the Luttinger liquid scheme are: no evidence of spin-charge separation at low frequencies, and the presence of broad, mostly dispersionless, features in $A(\mathbf{q}, \omega)$ in the frequency ranges of the order of one to few eVs [3,4] well below the Fermi level, and sometimes even below the band edges. While the absence of separated contributions of spin and charge collective degrees of freedom can be explained by the decrease of difference between corresponding boson phase velocities as coupling constants increase, the features at larger frequency scales cannot be simply interpreted even in this regime and are in fact beyond the scope of Luttinger liquid scheme which is formulated at low frequencies and becomes unreliable in the range of frequency cut-offs.

Putting aside the controversies related to the role of surface scatterings in photoemission measurements [5], two important points emerging from above considerations are to be emphasized. The first is that low frequency ARPES data are not in contradiction with Luttinger liquid scheme only for materials in which the medium and/or long-range interactions are strong. The second is that either the supplementary analysis of large energy scales or the inclusion of additional physical mechanisms [10] are needed to account for features observed in high frequency ARPES ranges.

## 2. PHYSICAL BACKGROUND OF THE RANDOM PHASE APPROXIMATION

Concentrating on the latter of above two points we consider one-dimensional band with three-dimensional Coulomb interaction and calculate its spectral properties at medium and large energies by utilizing the standard random phase approximation (RPA).

In order to justify such approach we first recall the well-known problem of electron self-energy for isotropic three-dimensional band, intensively studied within the jellium model in early days of many body theory. Lundqvist [11] and Hedin and Lundqvist [12] showed that then the changes in spectral density due to finite Coulomb interaction are mainly due to the collective plasmon mode, while incoherent electron-hole excitations introduce only nonessential corrections (like finite widths of quasiparticle peaks in the vicinity of Fermi edge). Plasmons drastically diminish the weight of these peaks (characteristically down to 50%), redistributing the spectral density towards the so-called *plasmaron* peaks and features in the plasmon frequency ($\omega_{pl}$) range above and below the Fermi edge.

It is expected that the influence of plasmon mode on the quasiparticles in the one-dimensional band is even more drastic, since its dispersion is then strongly anisotropic, covering in the long wavelength regime the whole frequency range $0 < \omega < \omega_{pl}$. Indeed, since this mode is just the collective mode in the charge sector of bosonized version of the jellium with linearized one-dimensional band dispersion [8,9,13], it gives the standard Luttinger liquid behavior at the Fermi edge with a finite anomalous dimension for $N(\omega)$. The bosonization technique for electron states however cannot be used in the wider energy range, i. e. roughly down to the bandwidth scale *t*. We relay to RPA in this range, noting that it perhaps gives qualitatively correct results even at small frequencies due to the cancellation of some non-RPA classes of diagrams for electron self-energy in this regime [14].

## 3. RESULTS AND CONCLUSIONS

The inverse electron propagator in RPA is given by

$$G^{-1}\left(k_{\parallel},\omega\right) = G_0^{-1}\left(k_{\parallel},\omega\right) - \frac{i}{2\pi N}\sum_{\mathbf{q}}\left[\int d\omega' \overline{V}(\mathbf{q},\omega') G_0\left(k_{\parallel}-q_{\parallel},\omega-\omega'\right) + i\pi V(\mathbf{q})\right] \quad (1)$$

with the bare propagator

$$G_0\left(k_{\parallel},\omega\right) = \frac{1-n(k_{\parallel})}{\omega - E_0(k_{\parallel}) + i\eta} + \frac{n(k_{\parallel})}{\omega - E_0(k_{\parallel}) - i\eta} \quad (2)$$

and the one-dimensional band dispersion $E_0(k_\parallel) = -2t(\cos k_\parallel a - \cos k_F a)$. Being concentrated on the long wavelength regime we take a simple form of bare Coulomb potential, $V(\mathbf{q}) = 4\pi e^2 / v_o |\mathbf{q}|^2$ where $v_0$ is the volume of the unit cell. The RPA enters through the dielectric function in the screened potential, $\overline{V}(\mathbf{q},\omega) = V(\mathbf{q})/\varepsilon_{RPA}(\mathbf{q},\omega)$, in which we retain only the plasmon contribution, $\varepsilon_{RPA}(\mathbf{q},\omega) = 1 - \Omega^2(\mathbf{q})/\omega^2$. The anisotropic plasmon dispersion is further simplified to its long wavelength form, $\Omega(\mathbf{q}) \approx \omega_{pl} |q_\parallel|/|\mathbf{q}|$, in order to facilitate the integrations in Eq.1. After the $\omega'$-integration Eq.1 reduces to

$$G^{-1}(k_\parallel,\omega) = G_0^{-1}(k_\parallel,\omega) - E_{ex}$$
$$-\frac{1}{2N}\sum_{\mathbf{q}} V(\mathbf{q})\Omega(\mathbf{q})\left[\frac{1-n(k_\parallel - q_\parallel)}{\omega - \Omega(\mathbf{q}) - E_0(k_\parallel - q_\parallel) + i\eta} + \frac{n(k_\parallel - q_\parallel)}{\omega + \Omega(\mathbf{q}) - E_0(k_\parallel - q_\parallel) - i\eta}\right] \quad (3)$$

where $E_{ex} = -\frac{1}{N}\sum_{\mathbf{q}} V(\mathbf{q}) n(k_\parallel - q_\parallel)$ is the exchange energy. The remaining $\mathbf{q}$-integration can be performed analytically, leading to the rather lengthy expressions for the propagator $G(k_\parallel,\omega)$ [15] omitted here for the lack of space.

The resulting spectral density $A(k_\parallel,\omega)$ for characteristic regimes of small and large plasmon frequency ($\omega_{pl}/t$ equal to 1 and 4 respectively) is shown in Fig.1. In both limits it has a form of a broad feature on the frequency scale with a smooth dispersion on the scale of longitudinal wavelength $k_\parallel$. The quasi-particle peaks are absent in the whole range of frequencies including the Fermi edge. As already announced, it is plausible to attribute this property to the characteristic anisotropic plasmon dispersion starting from acoustic low frequency contributions. In this respect the present results additionally amplify the early theoretical arguments pointing out a strong impact of plasmon collective modes on the renormalized electron spectrum. Recent analogous analysis showed that similar conclusions are valid also for (quasi) two-dimensional electron bands in layered structures [16].

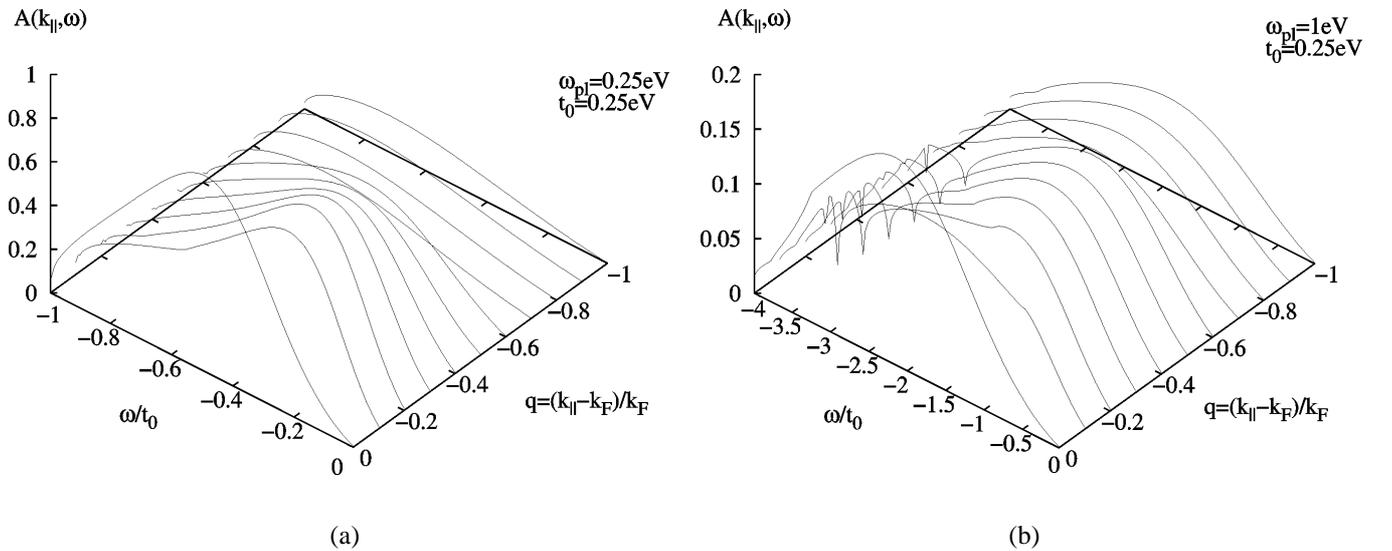

**Figure 1.** Spectral density $A(k_\parallel,\omega)$ for $\omega_{pl}/t = 1$ (a) and 4 (b).

In conclusion, the present analysis indicates that the notion of non Fermi liquid behavior of the spectral density follows already from the simple RPA for one-dimensional band with three-dimensional long-ranged Coulomb forces. It can be straightforwardly completed by the calculation of the exact value of anomalous dimension within the bosonization formalism. The comparison of two calculations will give insight into the validity (and applicability) of RPA in the studies of low energy scale in one-dimensional electron liquids. Another question to be answered within the present approach is that of the cross-over from the present non-Fermi liquid to the Fermi liquid regime as the transverse (inter-chain) bandwidth gradually increases. Both mentioned questions are under current investigations.

## Acknowledgements

We appreciate the communication of Prof. S. N. Artemenko (Russian Academy of Sciences, Moscow) on the work in Ref. [16]. The present work is supported by Croatian Ministry of Science and Technology, project no 0119251.